\documentclass[twocolumn,superscriptaddress,longbibliography,
	aps,prb,preprintnumbers,10pt]{revtex4-2}

\usepackage[normalem]{ulem}
\usepackage{graphicx}
\usepackage{bm}
\usepackage{color}
\usepackage{epsfig}
\usepackage{amsmath}
\usepackage{amssymb}

\usepackage[urlcolor=blue,colorlinks=true,citecolor=blue,linkcolor=blue,pdfstartview={FitH},bookmarks=false]{hyperref}

\usepackage{xcolor}
\graphicspath{{fig/}{./fig/}{.}}

\begin{document}
\title{Non-trivial phonon dynamics and significant electron-phonon coupling of the high frequency modes in a Dirac semimetal}
\author{Debasmita Swain}
\affiliation{Department of Physics, Indian Institute of Technology Kharagpur, Kharagpur 721302, India}

\author{Akash Dey}
\affiliation{School of Physical Science, National Institute of Science Education and Research, Bhubaneswar, 752050, India}
\affiliation{Homi Bhabha National Institute, Training School Complex, Anushakti Nagar, Mumbai 400094, India}

\author{Anushree Roy}
\affiliation{Department of Physics, Indian Institute of Technology Kharagpur, Kharagpur 721302, India}

\author{Kush Saha}
\email[e-mail: ]{kush.saha@niser.ac.in }
\affiliation{School of Physical Science, National Institute of Science Education and Research, Bhubaneswar, 752050, India}
\affiliation{Homi Bhabha National Institute, Training School Complex, Anushakti Nagar, Mumbai 400094, India}

\author{Sitikantha D. Das} 
\email[e-mail: ]{sitikantha.das@iitkgp.ac.in }
\affiliation{Department of Physics, Indian Institute of Technology Kharagpur, Kharagpur 721302, India}

\date{\today}

\begin{abstract}
Using finite temperature Raman spectroscopy, we investigate the electron-phonon interactions (EPI) and phonon-phonon scattering dynamics in the Dirac semimetal Cd$_3$As$_2$ in different frequency regimes. Strong softening of the Raman shifts below 200 K is observed for almost all the phonon modes with a marked deviation from the standard anharmonic behavior. The experimentally observed Raman linewidth seems to be captured well by a combination of EPI, relevant at low temperature (LT) and phonon-phonon scattering, which is predominant at high temperatures (HT), leading to an observable minima in the thermal evolution of the linewidth. While this feature is most prominently observed in the highest-frequency Raman mode ($\sim$196 cm$^{-1}$), its intensity gradually diminishes as the Raman frequency decreases. Computation of the electronic contribution to the phonon linewidth, for both the high and low frequency modes, from the phonon self-energy shows that it qualitatively mimics the experimental observations. It is found that phonon-induced interband scattering results in the presence of a maxima in phonon linewidth that crucially depends on the finiteness of the chemical potential. 
\end{abstract}

\maketitle


Dirac semimetals exhibit a variety of interesting phases and properties such as giant diamagnetism with very high electron mobilities~\cite{PhysRevB.81.195431}, large HT quantum magnetoresistances~\cite{PhysRevB.58.2788, PhysRevLett.106.156808}, pressure induced superconductivity~\cite{He2016} and many more. While there exists a plethora of physical phenomenon where EPI can be taken as a perturbation under the adiabatic Born Oppenheimer approximation, this approach fails in certain cases, such as moderately correlated systems, graphene, and 2D Dirac crystals~\cite{Pisana2007}. Transport and optical properties reveal Kohn anomalies in these systems, where electron and phonon dynamics are intertwined~\cite{PhysRevLett.93.185503}.Dimensional reduction in the form of thin films exhibit weak antilocalisation in magnetoresistance measurements, which is a manifestation of chiral anomaly as expected in Dirac semimetals~\cite{Li2015}. Unambiguous, Aharanov-Bohm oscillations in Cd$_3$As$_2$ nanowires provide evidence of non-trivial Fermi arc like surface states  whereas such effects are rather difficult to observe in the bulk form due to volume dominated conduction bands~\cite{Zhao2016,Wang2016}. This shows that electrical transport behaviour is remarkably dependent on the dimensionality of the underlying lattice structure via the bulk boundary correspondence. Understanding how linearly dispersing massless Dirac fermions interact with lattice vibrations is, therefore, crucial for explaining these phases.  In view of that, we aim to explore graphene-like physics in Cd$_3$As$_2$~\cite{PhysRevLett.128.045901},  one of the very few stable 3D Dirac crystals under ambient condition~\cite{doi:10.1126/science.1245085,Xu2017,Liu2014AST}.

Despite the topologically protected states, transport in Dirac materials (DMs) is affected by various scatterers, such as charged impurity, magnetic impurity, static disorder, and phonons, amongst others. Since phonons are ubiquitous in materials, the electron-phonon and phonon-phonon scatterings are inevitably present even in pristine crystals to modify material properties. Thus the study of phonon dynamics is crucial for understanding transport as well as thermal properties in DMs. It is already known that phonons can modify band topology in Dirac insulators~\cite{PhysRevLett.110.046402}. Moreover, some modes are argued to carry signature of topology in Dirac insulators. Although there are several recent theoretical studies on phonons in DMs\cite{garate_PRL2013,saha_PRB2014,saha_PRL2015,antonio_PRL2016,vanderbilt_PRL2016,PhysRevLett.119.107401,PhysRevLett.125.146402,PhysRevB.109.144304}, the experimental evidence of phonon modes and related phenomena is limited to only a few DMs.  For example, the decay mechanism of both infrared active phonons \cite{Xu2017} and Raman active phonons \cite{PhysRevB.100.220301} in mono-arsenides Weyl semimetals such as TaAs and NbAs have been studied in recent years. Also, the observation of very few Raman modes in single crystals of Cd$_3$As$_2$ was reported so far without subsequent in-depth investigations~\cite{https://doi.org/10.1002/jrs.1250150215, Gupta_2017}. This further led to the study of phonon dynamics in Cd$_3$As$_2$ \cite{Sharafeev.Lemmens.17}, and a complex interplay between electronic and phonon degrees of freedom has been discussed. Notably, this specific study highlights the involvement of lower frequency phonons associated with the vibration of Cd ions. The observed anomalous anharmonicities in the Raman shift and line width of these phonons stem from their coupling to Dirac states near the Fermi energy. On the contrary, the high-frequency phonons, attributed to the vibration of As ions are suggested to be unrelated to Dirac states. In fact,  the high-frequency Raman study of Cd$_3$As$_2$ is yet to be investigated in detail. 

In view of this, we revisit phonon dynamics of Cd$_3$As$_2$ through temperature-dependent Raman measurement across a wide temperature range and in the various frequency regimes. In Cd$_3$As$_2$, we have observed an anomalous behaviour of Raman shift and linewidth with temperature for both low and high frequency (HF) phonon modes. Of particular interest is the discrepancy in the magnitude of this anomalous behavior between the low and HF phonon mode regimes. To interpret the mechanism driving this phenomenon, we analyse electron-phonon coupling (EPC) utilizing the {\it k.p} model calculation. Specifically, our study delves into the correction to phonon self-energy resulting from EPC interactions. Through this analysis, we gain valuable insights into the role of electronic transitions within Dirac bands, revealing the underlying mechanism behind the observed behavior.

 Time reversal and crystalline symmetry puts significant constraints on the band structure which has been discussed in detail elsewhere~\cite{Liu2014AST}. The origin of linear electronic bands is related to the band crossing that occurs around gamma point between the $5s$ bands from Cd ions with a positive slope and $3p$ bands of As ions with a negative slope~\cite{PhysRevB.102.165115} resulting from band inversion brought about by the presence of significant spin-orbit coupling. While the relative contribution of the cationic $s$ states and the anionic $p$ states at the Fermi level might result in variations in the effective low energy models relevant for Cd$_3$As$_2$, surely the contribution of both is important. Thus, the high energy Raman modes, which are primarily due to As vibrations, should also carry signatures of the nontrivial electronic structure.

\begin{figure}[h!]
	\centering
	\includegraphics[width=\linewidth]{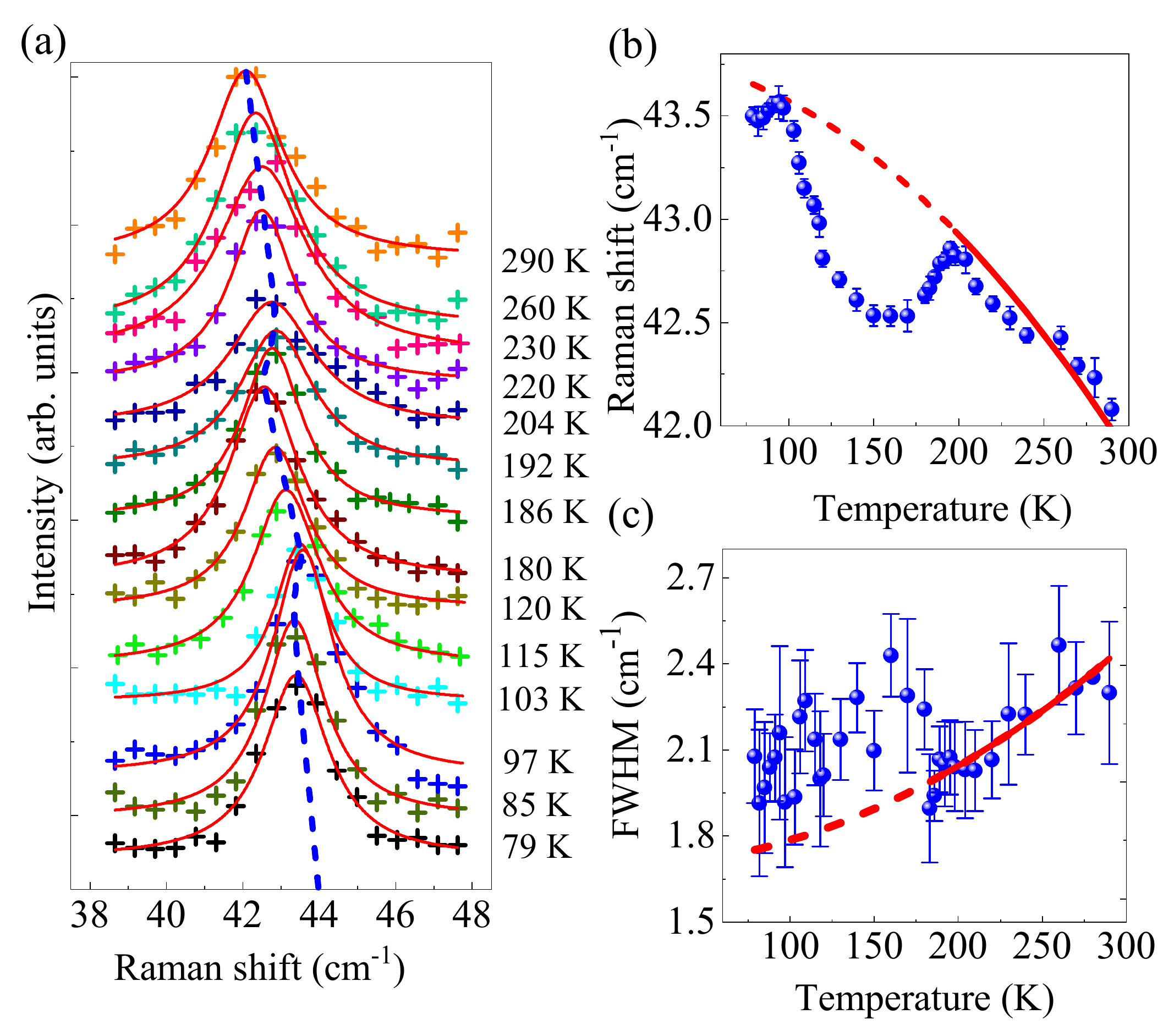}
	\caption{(Color by electronic media) (a)  Characteristic Raman spectra of Cd$_3$As$_2$ between 80 K to 300 K (given by + symbols), focusing on the peak appearing around 42 cm$^{-1}$ with Lorentz fitting (given by red solid line). The variation of peak position with temperature is marked by blue dashed line. Temperature evolution of (b) Raman shift with anharmonicity curve fitting (3-phonon(3ph) and 4-phonon(4ph)) by equation SE1 and (c) FWHM of the corresponding peak. Red solid lines represent HT fits till 200 K, and dashed lines are extrapolated down to 100 K.}
	\label{Raman2}
\end{figure}

 Figure \ref{Raman2} (a) shows the temperature dependent low frequency (LF) Raman spectra taken on the single crystal of centrosymmetric compound Cd$_3$As$_2$ on the \{1 1 2\} plane (for more details about basic characterisation and Raman data, see supplementary material (SM)~\footnote{See Supplemental Material at [URL will be inserted by publisher] for additional experimental results presenting sample characterisation and temperature dependent Raman spectra, and theoretical results. See also references \cite{Sankar2015,Ali2014,Giustino,Novko} in the SM}). Initially, we look at the 42 cm$^{-1}$ Raman mode that belongs to the low-frequency regime (along with 27 cm$^{-1}$ and 35 cm$^{-1}$) and compare with related modes with higher frequency, it is well separated from others which are convoluted with relatively close neighbouring peaks. 
The temperature evolution of Raman shift and full width half maxima (FWHM), acquired from the fitting, are presented in Fig.\ref{Raman2} (b) and (c), respectively. Notably, from the plots, a significant anomaly is observed below 200 K, deviating from the anticipated anharmonicity behavior of optical phonons (given as red dashed line in Fig.~\ref{Raman2}). This is attributed to the intricate interplay between phonon and electronic degrees of freedom of the quasielastic background, as also discussed by Sharafeev et al.~\cite{Sharafeev.Lemmens.17}. The temperature evolution of the Raman shift of other LF modes is given in the supplementary information (see Fig. SF3, SF4), and it is important to note that all the modes within the scope of our study exhibit deviations from the expected anharmonic behavior below 200 K.

\begin{figure}[h!]
	\centering
	\includegraphics[width=\linewidth]{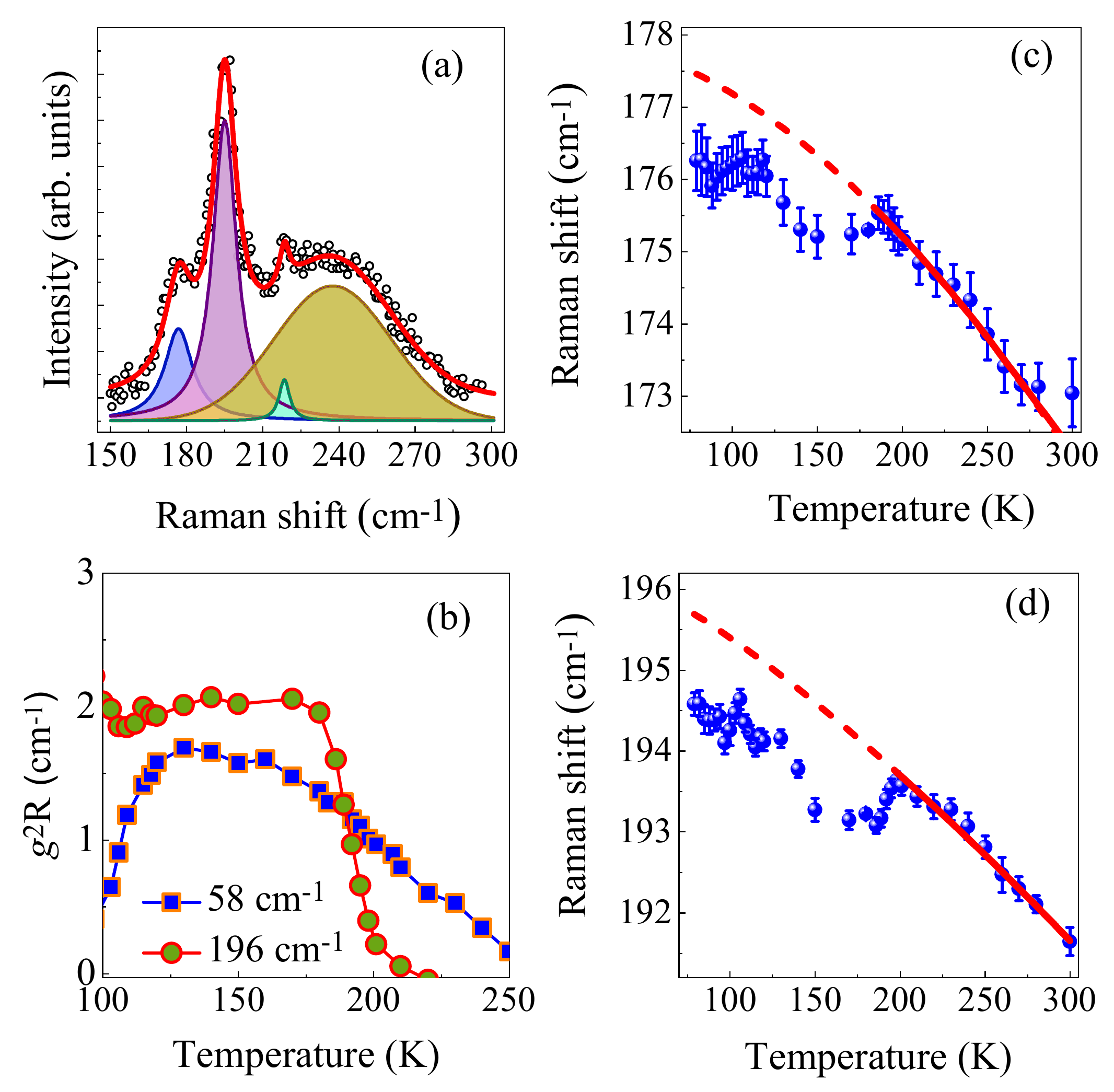}
	\caption{(Color by electronic media)  (a) Background subtracted deconvoluted high frequency Raman spectra of Cd$_3$As$_2$ (black hollow circles) at 80 K with the three peaks (shaded with blue, magenta, and cyan colours) are fitted with Lorentzian profile and the broad maxima (olive green) with Gaussian profile (red solid line: cumulative fit). (b) $g^2R$ plotted within a temperature range between 100-250 K for the Raman modes at 58 cm$^{-1}$ and 198 cm$^{-1}.$ 
    Panels (c) and (d) show the corresponding Raman shift of 177 cm$^{-1}$ and 196 cm$^{-1}$ peaks respectively (blue solid balls). The Raman shift of the modes are fitted by 3ph and 4ph anharmonicity equation~\cite{balkanski.wallis.83} (red solid lines) and extrapolated ones (red dashed line).}
	\label{Raman3}
\end{figure}

Figure \ref{Raman3}(a) presents background-subtracted deconvoluted high frequency Raman spectra of Cd$_3$As$_2$ (depicted as black hollow circles) at 80 K. Evidently, three distinct phonon peaks are very prominent and upon comparing with the available literature~\cite{Sharafeev.Lemmens.17, Gupta_2017}, the modes around 175 cm$^{-1}$ and 196 cm$^{-1}$ are identified as B$_{1g}$ symmetry with the weak mode around 220 cm$^{-1}$ as A$_{1g}$ symmetry. These peaks are accompanied by a broad Gaussian maximum centered around 236 cm$^{-1}$. The peak positions of the B$_{1g}$ Raman modes obtained from the fitting are plotted in Fig. \ref{Raman3}(c) and (d), respectively, as a function of temperature. Red solid lines (dashed lines are extrapolated ones) are the anharmonic curve fitting to the data as per the anharmonicity equation SE1, given in SM~\cite{Note1}. Intriguingly, from the plots, a discernible change in slope is observed, deviating from the expected anharmonicity behavior of optical phonons. 

The broad Gaussian maxima observed within the frequency range between 225 to 300 cm$^{-1}$ exhibits an FWHM of $\sim$ 35 cm$^{-1}$. Such broad peaks have recurrently been observed in other topological systems~\cite{PhysRevB.84.195118, article, PhysRevB.95.245406}, although the origin of this mode is yet to be clearly understood. Similar to our observed feature, Sharafeev et al.~\cite{Sharafeev.Lemmens.17} has also noticed such broad Gaussian maxima in Cd$_3$As$_2$, almost in the comparable frequency range, and the broad maxima has been attributed to the result of electronic Raman scattering on the surface Dirac states, based on the studies of Rashba semimetals~\cite{article} and others~\cite{PhysRevB.84.195118}. Signature of such electronic Raman scattering on the surface states has been found below 100 cm$^{-1}$ for both pristine and Cu doped Bi$_2$Se$_3$~\cite{PhysRevB.84.195118}, while our observed value as well as the reported case~\cite{Sharafeev.Lemmens.17} is at HF. In contrast, extensive studies on topological insulator Bi$_2$Se$_3$ have revealed a similar broad feature near a frequency range of our study, which is identified as stemming from two-phonon excitation. A comprehensive series of doping-dependent studies have distinguished this mode from the surface-originated electronic modes~\cite{PhysRevB.95.245406}. 

EPI modifies the electronic spectral function, and for low coupling regimes, it is normally studied within the Allen Heine Cardona formalism~\cite{PhysRevB.105.245120}. In our case, we take a simplistic approach to understand the observed deviation of the Raman frequency shift from the behaviour expected from the Klemmens model of anharmonic decay~\cite{PhysRev.148.845}. A significant deviation is observed in almost all the Raman modes below 200 K, as described earlier. Such softening of the phonon mode frequencies can result from a variety of reasons like spin-phonon coupling, interaction with magnons or EPC. As Cd$_3$As$_2$ has no magnetic order, this anomaly must, therefore, result from EPC. At the phenomenological level, EPC ($g$ is the EPC constant) would result in the modification of the observed phonon frequency $\omega_{obs}(T)$  as $ \omega_{obs}(T) = \sqrt{\omega_{anh}(T)^2 - g^2R(\omega,T)\omega_{anh}(T)}$, where 
$\omega_{anh}(T)$ is the anharmonic contribution to the bare Raman mode frequency $\omega$ ($\omega_{anh}(T)$ is given by SE1) and
$R(\omega,T)$ is the real part of the electric susceptibility of the mobile charge carriers. Here $\omega_{anh}(T)$ is estimated from the value of the Raman shift found by extrapolating the HT data~\cite{PhysRevB.101.245431}.The thermal variation of  $g^2R$, for various Raman modes, is shown in the range between 100 - 200 K in Fig. 2(b). It is seen that for both the low frequency 58 $cm^{-1}$ as well as the high frequency 196 $cm^{-1}$ mode, the value of  $g^2R$ has a relatively small variation in magnitude between 1.5 to 2.0, an observation which is consistent with the fact that it would be unphysical to expect a large variation of $g^2R$ across different Raman modes (see SF5). It is interesting to note that the $g^2R$ generally increases with frequency. However, it is not known if $R$ and $g$ individually or together increase with the frequency~\cite{PhysRevB.101.245431}. If it is assumed that the variation of $R$ is relatively small with frequency, then that would imply $g$ increases with Raman mode frequency. In general, information about band topology is much less reflected in the Raman frequency shift than in the linewidth due to which we look in detail at the thermal evolution of phonon linewidth subsequently.

\begin{figure*}
\includegraphics[width=\linewidth]{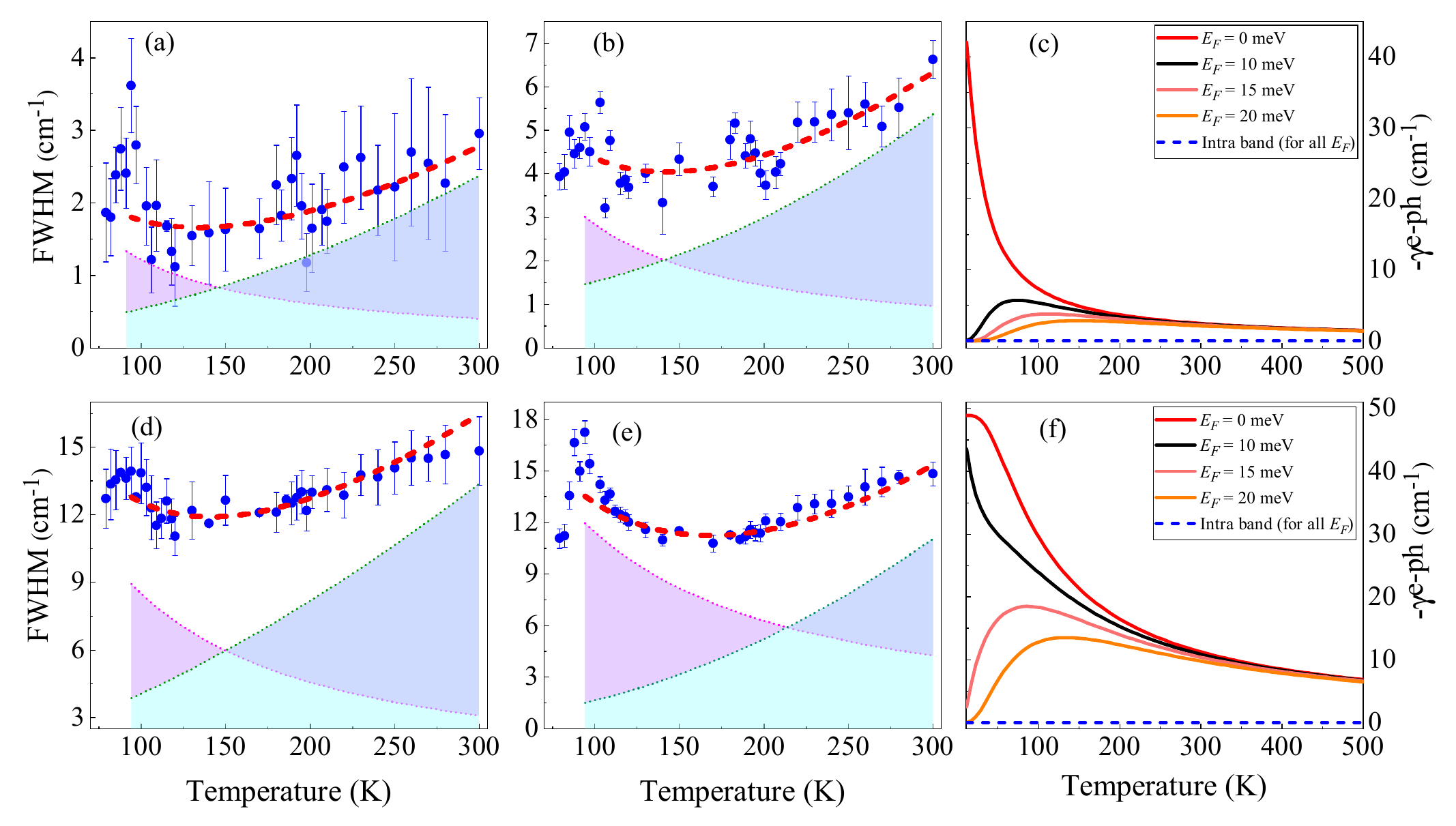}
\caption{(a) FWHM plot of Raman peak at 27 $cm^{-1}$, (b) 58 $cm^{-1}$, (d) 177 $cm^{-1}$ and (e) 196 $cm^{-1}$, are given by blue solid circles. The red dashed line through the data represents the total e-ph (for $\mu$ =0) and anharmonicity fit (3ph and 4ph). The magenta and green dashed lines are the individual e-ph equation and (3ph,4ph) anharmonicity equations respectively, generated by taking the above fitting parameters from the total fit.(c) Plot of phonon linewidth as a function of temperature with phonon frequency $\omega=42 cm^{-1}$ and phonon momentum $q=0\AA^{-1}$. Here the solid lines represent interband contribution at different Fermi energies and the dashed line is for intraband contribution for all the Fermi energies. (f) Same as (c) for the phonon frequency $\omega=195 cm^{-1}$.}
\label{fig:Interband}
\end{figure*} 

The FWHM plots of the LF and HF Raman modes are given in figure \ref{fig:Interband}(a-b), and \ref{fig:Interband}(d-e), respectively. The linewidth of both the LF and HF modes exhibit a significant dependence on temperature. Following Klemmens model, considering both 3ph and 4ph anharmonic decay process which involves the decay of an optical phonon into two or three acoustic phonons, respectively, the temperature dependence of the Raman linewidth ($\Gamma (T)$) is captured by the following equation~\cite{PhysRev.148.845}:

\begin{eqnarray}
\label{anh2}
\Gamma_{anh}(T) = P_o &+& P_1\left( 1+\frac{2}{e^{\frac{y}{2}}-1}\right)  \\
\nonumber &+&   P_2\left( 1+\frac{3}{e^{\frac{y}{3}}-1}+\frac{3}{(e^{\frac{y}{3}}-1)^2}\right),
\end{eqnarray}
where $y = \frac{\hbar\omega}{k_BT}$. $P_0$, $P_1$ and $P_2$ are anharmonic constants.
Figure \ref{fig:Interband} shows that the FWHM data at HTs align well with the anharmonicity model. However, at temperatures below $\sim$ 200 K, the behavior diverges from the predictions of equation \ref{anh2}. This discrepancy suggests a possibility that optical phonons may be decaying into fermionic excitations via interband or intraband transitions close to the Fermi level at lower temperatures~\cite{PhysRevB.73.155426, PhysRevLett.99.176802, PhysRevX.11.011017}, which is important particularly in case of semimetals. \\

\textit{Theoretical Study: Phonon self-energy}. To gain insights into the experimental outcomes, we focus on the scattering of electrons from phonons, which in turn can shift phonon frequencies and contribute to the phonon linewidth. Thus to find the effect of EPI, we compute phonon self-energy~\cite{saha_PRL2015, heid15}:
\begin{equation}
\label{eq:pi}
\Pi_\lambda({\bf q},\omega_{{\bf q}\lambda})=\frac{1}{\cal V}\sum_{{\bf k} nn'} \frac{|g^\lambda_{nn'}({\bf k,q})|^2\left(f_{{\bf k}n}-f_{{\bf k-q}n'}\right)}
{E_{{\bf k}n}-E_{{\bf k-q}n'}-\omega_{{\bf q}\lambda}-i 0^+},
\end{equation}
where ${\cal V}$ is the sample volume, $\lambda$ labels different phonon modes, ${\bf q}$ is the phonon momentum, $\omega_{{\bf q}\lambda}$ is the bare phonon frequency and  $f_{{\bf k}n}$ is the fermion occupation number for the state $|u_{{\bf k}n}\rangle$ with a Fermi energy $E_F$. Here $g^\lambda_{nn'}({\bf k,q})=\langle u_{{\bf k}n} | {\hat g}^\lambda ({\bf q})|u_{{\bf k-q} n'}\rangle$, where ${\hat g}^\lambda ({\bf q})$ is the electron-phonon vertex operator.${\hat g}^\lambda ({\bf q})$ can take the matrix form corresponding to either even or odd parity phonon modes, however, in the present case we have considered the even parity phonon mode (refer to Fig.SF7 in the SM~\cite{Note1}). The Raman shift of the phonon mode is given by the real part of phonon self-energy, which is discussed along with the analysis in the SM~\cite{Note1}. It is evident that while there is some level of qualitative agreement, the variation in the calculated values are two orders of magnitude smaller than that obtained experimentally. This quantitative disagreement possibly indicated that the Raman shift calculated by including only electron-phonon coupling may not be able to completely capture the scenario and further investigation is necessary. The linewidth of phonon mode is given by the imaginary part of phonon self-energy, as per equation SE3 in SM~\cite{Note1}. This equation can be simplified upon considering a linearized band dispersion around Fermi energy and, hence, the temperature dependence of linewidth can be quantitatively expressed in terms of the difference of occupation of electronic states below and above Fermi energy as~\cite{PhysRevB.73.155426, PhysRevLett.99.176802}, 
\begin{equation}
\label{eq:Fermi}
\gamma_{e-ph} = A  \left [f \left ({\it -B/T}\right)- f \left ({\it B/T}\right)\right],
\end{equation}
 where A is the EPC constant at T = 0 K, {\it B} = $\frac{\hbar \omega}{2k_B}$, {\it f(B/T)} is the Fermi-Dirac distribution function. It is to be noted that the finite chemical potential is not taken into account in equation~\ref{eq:Fermi}.

 In the present scenario, to fit the Raman linewidth with temperature, we use an expression with both anharmonicity (as given in eqn.~\ref{anh2}) and EPC equation (see eqn.~\ref{eq:Fermi}) as, $\Gamma$ = $\Gamma_{anh}$ + $\gamma_{e-ph}$.  
From the plots (see Fig.~\ref{fig:Interband} (a-e), the LT data is fitted well by the EPC equation, as evident from the individual contribution equation plotted as magenta curve, whereas the data above 200 K is well explained by the 3ph and 4ph anharmonicity equation, given by green curve. If the area between the individual EPC curve and anharmonicity curve at LT, can be taken as a quantitative measure of the EPC contribution, then it is interesting to observe that both the LF and HF modes are dominated by EPC at LT with a decreasing trend of EPC constant {\it A} (from equation~\ref{eq:Fermi}) from HF to LF (see Fig.SF6). This, along with the observation from Raman shift, indicates that EPC is dominant at HF range with a comparatively weak signature at LF range. Moving to the HT data, the ascending trend of linewidth is accurately explained by both  3ph and 4ph scattering terms as, including 4ph term leads to a considerable change to phonon self-energy~\cite{PhysRevLett.128.045901}. It is worth pointing out that all modes show a minima in the thermal evolution of linewidth at around 200 K. Terahertz experiments on Cd$_3$As$_2$ have confirmed that a phonon bottleneck effect is observed with the interband relaxation time increasing with temperature below 220 K and remaining approximately constant above it till 300 K. It is quite possible that this cross over from a predominantly electron - phonon coupling dominated regime to a phonon - phonon interaction mediated regime in the Raman linewidth at around 200 K is also related to that observed change in carrier dynamics around 220 K~\cite{PhysRevB.106.155137}.

The temperature dependency of the phonon line width is governed by the Fermi functions in Eq. SE3 and it depends on the doping level (or the Fermi energy). In Fig.~\ref{fig:Interband}(c) and ~\ref{fig:Interband}(f), we show the variation of intra and interband phonon-linewidth, $\gamma$ with temperature for $q\simeq 0$ phonons. To make our theoretical results relevant to experimental outcomes, we consider only two frequencies of phonons,  $\omega=42$ cm$^{-1}$ and $\omega=195$ cm$^{-1}$ associated with the Raman peaks in Fig.~\ref{Raman2} and Fig.~\ref{Raman3}, respectively. Notice that in both cases, $\gamma$ is dominated by interband contribution (see SM~\cite{Note1}). We further find that $\gamma$ overall reduces with temperature but may show peaks at some particular temperature depending on the doping of the bands or in other words, on the Fermi energy. For simplicity, we focus on the $\gamma$ at HF and comment on the low-frequency case.

At $T=0$ K and $E_F = 0$ meV, the valence band is fully occupied. Therefore, only interband transitions from the valence band ($n'=v$) to the conduction band ($n=c$) are possible, contributing maximally. As the temperature increases, the term $f_{{\bf k}c}(T) - f_{{\bf k}v}(T)$ in Eq.(SE3) decreases due to the rise of electrons' population in the conduction band. Accordingly, the transition probability decreases. This is illustrated in Fig. \ref{fig:Interband}.
As $E_F$ is raised (e.g., $E_F=20$ meV), the conduction band population increases even at $T=0$ K, reducing $\gamma$ to nearly zero as evident from Fig. \ref{fig:Interband}. As the temperature increases, electrons start populating higher momentum states in the conduction band, causing the matrix elements to increase and reach a peak value. However, with further temperature increases, the delta function in Eq. SE3 limits the momentum states available for electron scattering. As a result, the contribution to $\gamma$ from the matrix elements decreases, leading to an overall reduction in $\gamma$ with temperature (fig- SF7 for more clarification). Interestingly, the peak feature qualitatively appears approximately at the same temperature obtained in experiments (in fig-~\ref{fig:Interband}). We note that the explanation above can easily be extended to the low-frequency phonons (see Fig. \ref{fig:Interband} c). 

Overall, an attempt has been made to explain the Raman data using a phenomenological model of a Dirac semimetal in the limit of low-energy excitations. While there is a qualitative agreement, there are several aspects where deviations are observed. It is important to point out that, while there is definite evidence of softening of the Raman frequencies at temperatures less than 200 K, the estimated values of $g^2R $ obtained from the deviations do depend on the parameters of the anharmonic fitting function. For instance, forcing a fit to converge to the Raman shift values at around 100 K (Fig~\ref{Raman3}) would lead to a $\omega_0$ of $\sim$ 194 cm$^{-1}$ with reduced, albeit non zero values, of electron-phonon coupling. This would however, imply a resurgence of the standard anharmonic behaviour at low temperature, which is not likely to be consistent with the thermal evolution of the phonon linewidth in this temperature range, thereby in turn justifying the analysis as discussed earlier. It is also worth noting that for Cd$_3$As$_2$, the typical chemical potential estimated from bulk thermodynamic measurements varies between 150 - 200 meV~\cite{Liu2014AST}, which has not been taken into account for estimating the values of $g$ ($A$, in case of FWHM) from the observed phonon linewidths using the eqn.~\ref{eq:Fermi}. The theoretically calculated phonon linewidths have a significant quantitative dependency on the chemical potential as discussed previously. While it agrees with the experimental observation qualitatively, it can be seen that the peak structure around 100 K is much more sharper in the experimental data than that which is obtained theoretically. This discrepancy might be a manifestation of the fact that the asymmetric nature of the pair of Dirac cones, with realistic tilting parameters relevant for Cd$_3$As$_2$ has not been considered. It is known that there is an increase in the density of states and decrease in the chemical potential due to the tilting of the bands~\cite{PhysRevB.108.L161401}.  Furthermore, while it is now accepted that at least in the low energy regime, the physics of  Cd$_3$As$_2$ is governed by Dirac model while at higher energies, massless Kane electrons might be prevalent with the Bodnar model~\cite{PhysRevLett.117.136401} being the relevant microscopic model, there are experimental evidences which indicate that this issue has not yet been completely settled. In this respect, we show that while we can explain certain aspects of our experimental findings through a minimalistic model, perhaps more detailed experimental and theoretical considerations are required.

\textit{Summary}. In conclusion, in this work, we have investigated the temperature evolution of the Raman active modes of Cd$_3$As$_2$, specifically studying the shift in phonon frequency and linewidth. We show that below 200 K, there is a significant redshift of the Raman frequency from the expected anharmonic behaviour for both the LF as well as the HF modes which indicates that the phonon modes linked to both the cadmium as well as the arsenic atoms are coupled to the Dirac electrons, contrary to what was reported earlier. The thermal evolution of the linewidth shows a pronounced minima stemming from the collective contributions of the electron-phonon and the phonon-phonon scattering terms. Interestingly, the strength of the EPI diminishes progressively with the decrease in Raman mode frequency, in concomitance with linearly dispersing bands. In HT phonon-phonon interaction regime, it was observed that four phonon scattering term is essential to successfully explain the experimental data. By calculating the phonon self-energy using a four band Dirac model Hamiltonian, we show that interband scattering results in an electron-phonon linewidth variation similar to what is observed experimentally. Our work highlights the importance of EPI and phonon-phonon interaction in Dirac semimetals and paves the path for further investigation.

S.D.D. also acknowledges the financial support of ISIRD, IIT Kharagpur and SERB through grant No. ECR/2017/003083. 
A.R. and SDD acknowledge the financial support of SERB India through grant No. CRG/2021/000718. SDD acknowledges the help of P P Jana and S K Kuila for crystal orientation. KS acknowledges funding from the Science and Engineering Research Board (SERB) under SERB-MATRICS Grant No. MTR/2023/000743. 

\bibliography{biblio.bib}

\end{document}